\documentclass[sigconf,nonacm]{acmart}

\copyrightyear{2022}
\acmYear{2022}
\setcopyright{rightsretained}
\acmConference[ASIA CCS '22]{Proceedings of the 2022 ACM Asia
Conference on Computer and Communications Security}{May 30-June 3,
2022}{Nagasaki, Japan}
\acmBooktitle{Proceedings of the 2022 ACM Asia Conference on
Computer and Communications Security (ASIA CCS '22), May 30-June 3,
2022, Nagasaki, Japan}\acmDOI{10.1145/3488932.3527292}
\acmISBN{978-1-4503-9140-5/22/05}

\usepackage{titlesec}
\titlespacing*{\section}
{0pt}{0.2cm}{0.2cm}

\begin{document}
\fancyhead{}

\title{Email Summarization to Assist Users in Phishing Identification}

\author{Amir Kashapov}
\email{akas0005@student.monash.edu}
\affiliation{%
  \institution{Monash University}
  \country{Australia}
}

\author{Tingmin Wu}
\email{tina.wu1@monash.edu}
\affiliation{%
  \institution{Monash University, CSIRO's Data61}
  \country{Australia}
}

\author{Alsharif Abuadbba}
\email{sharif.abuadbba@data61.csiro.au}
\affiliation{%
 \institution{CSIRO's Data61}
 \country{Australia}
 }
 
\author{Carsten Rudolph}
\email{carsten.rudolph@monash.edu}
\affiliation{%
  \institution{Monash University}
  \country{Australia}
}

\begin{abstract}

Cyber-phishing attacks recently became more precise, targeted, and tailored by training data to activate only in the presence of specific information or cues. They are adaptable to a much greater extent than traditional phishing detection. Hence, automated detection systems cannot always be 100\% accurate, increasing the uncertainty around expected behavior when faced with a potential phishing email. On the other hand, human-centric defence approaches focus extensively on user training but face the difficulty of keeping users up to date with continuously emerging patterns. Therefore, advances in analyzing the content of an email in novel ways along with summarizing the most pertinent content to the recipients of emails is a prospective gateway to furthering how to combat these threats. Addressing this gap, this work leverages transformer-based machine learning to (i) analyze prospective psychological triggers, to (ii) detect possible malicious intent, and to (iii) create representative summaries of emails. We then amalgamate this information and present it to the user to allow them to (i) easily decide whether the email is ``phishy'' and (ii) self-learn advanced malicious patterns.


\end{abstract}


\keywords{Phishing, Email, Machine Learning, Summarization}


\maketitle

\section{Introduction}
Phishing is a type of cyber-attack whereby criminals design seemingly authentic emails with the intent of tricking users into giving up private, confidential, and/or sensitive information such as login passwords and financial data. Artificial Intelligence-based automated phishing detection methods are limited as Machine Learning (ML) models can only identify the malicious patterns they have been trained on and detection by heuristics-based techniques produce high false positive rates~\cite{basit2021comprehensive}. In today's highly connected world, users spend considerable time sending and reading both work and personal emails, and it is thus hard for users to pay close attention to every email. Consequently, they can fall victim to advanced deceptive phishing techniques that bypass security filters. 

Recent work shifts the focus from automated phishing detection to detection support to assist users in making their own judgements~\cite{althobaiti2021don}, 
since automated approaches are not always 100\% accurate and real-time support such as warnings can effectively change risky behaviors. Presenting users with security indicators' information enables human strength in capturing abnormal behaviors, such as contextual awareness.
A human-centric solution using autonomous ML agents to aid judgment can therefore be a crucial step in the right direction. Despite security warnings and modern means of educating users about phishing, current detection support and training still cannot deal with sophisticated malicious emails in real time. Therefore, condensing important email information relevant to phishing and rationalising human efforts are important objectives to deal with new types of attacks.

Most of the existing works on phishing detection support heavily rely on the legitimacy of URLs~\cite{alsharnouby2015phishing,althobaiti2021don}.
However, explanations of URL features are easy to understand by general users, e.g., website rank, hostname, and domain popularity. 
More importantly, phishing attacks have evolved to escape URL detection by leveraging social engineering. In particular, criminals can generate links through third-party services or a plaintext with a response request~\cite{almashor2021characterizing}. Motivated by that, we aim to address the challenges of improving the readability and effectiveness of generated information on phishing emails that URL-based detection methods cannot detect. 

Therefore, the objective of this work is to design and develop a human-centric notification system based on both emails and their contexts using Natural Language Processing (NLP) methods to help users accurately identify phishing attempts. The intention is to outline useful information based only on the email and the contextual knowledge that machines can utilize - to be precise, this includes the summary of an email, the emotions associated with an email and the intents of the
sender that may be relevant to evoking  self-sabotaging actions.

\subsection{Contributions}

We present a novel, human-centric mechanism to combat one of the most prevalent cybercrimes of the world today - email phishing. Our contributions to that end can be summarized as follows:

\begin{enumerate}
  \item We propose a system to pipeline emails through ML models and to generate a human-centric summary report on useful criteria for the user to identify phishing. 
  \item We leverage and investigate the utility of the summarization capability of the latest transformer-based ML models on emails for phishing analysis.
  \item We leverage and investigate the utility of the emotion classification capability of transformer-based ML models on emails in the context of ``cognitive triaging'' to detect potential psychological triggers.
  \item We leverage and investigate the utility of the intent analysis capability of transformer-based ML models on emails to determine whether malicious intent can be accurately ``cherry-picked''.
\end{enumerate}

\section{System Design}
We identified three critical components in particular as utilities of state-of-the-art ML models that could indicate potential phishing emails, and, as a result, we build our system to combine them. The system design thus incorporates three fundamental pipelines: extractive summarization, emotion classification, and intent analysis. We next explain those three pipelines.\\

\begin{figure}[h]
        \includegraphics[width=1\columnwidth]{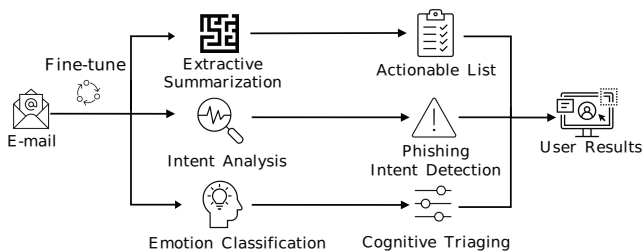}
        \caption{\label{architecture.png}System Design Overview
        }
\end{figure} 

\textbf{Extractive Summarization}. 
Phishing emails can be overwhelmingly long to deceive the recipients to overlook the phishing components and to focus on urgency and/or emotions. 
To emphasize the most important context, 
we use the summarization capability of modern ML models. Informing the user of the most important contents of the email and of the potentially desired actionable items evoked by the sender could allow them to examine the primary purpose of the email without any superfluous or distracting information clouding their judgement. For instance, if a small part of the email evokes a potential victim to click on a harmful link whereas the others are merely warnings of what would occur if they don’t, an ideal summarization branch/pipeline would immediately point this out as as irregular as part of the user results.

\textbf{Emotion Classification}. 
The multi-class emotion classification branch is inspired by the brilliant paper on cognitive triaging~\cite{van2019cognitive}. There are six possible ``cognitive triggers'' we detect in phishing emails  - \textit{Reciprocity, Consistency, Social Proof, Authority, Liking, and Scarcity}~\cite{van2019cognitive}. The definition of these terms are, respectively, as follows: \textit{Reciprocity} - Tendency to feel obliged to repay favors from others. \textit{Consistency} - Tendency to behave in a way consistent with past decisions and behaviors.  \textit{Social Proof} - Tendency to reference the behavior of others to guide one's own actions. \textit{Authority} - Tendency to obey people in authoritative positions. \textit{Liking} - Preference for saying ``yes'' to the requests of people one knows and likes. \textit{Scarcity} - Tendency to assign more value to items and opportunities when their availability is limited \cite{van2019cognitive}.  


\textbf{Intent Analysis}. 
This analysis filters and detects potentially malicious intent in the form of actionable and identifiable items that can be detected through the use of ML models. 
The aim of this branch is that a recipient gets a clearer picture as to whether an email is harmful when highlighted and contextualized suspicious portions of an email are taken into consideration with information from the other branches. Highlighting potential phishing intent could also serve to train recipients to develop a ``feel'' for what phishing intents and signals look like.


\section{Preliminary results}

\subsection{Data collection and pre-processing}

The first experimental setup step was to compile email datasets which include the Cambridge phishing dataset, the Cornell phish bowl, the Enron email corpus of benign emails, the millersmiles.co.uk phishing dataset, and the Nazario phishing dataset. The above were chosen as they were collected by reputable researchers to create an accurate representation of phishing contexts or of benign correspondence in emails - in fact, all datasets in question compile real-life emails. The corpora contain 41446, 1757, 252721, 33080, and 946 emails respectively, and were all utilized to create random experiment samples of varying sizes in a variety of contexts - early evaluation sample sizes numbered 500 or more emails followed by final evaluation samples numbering 5000 or more for all branches.

For all three branches, pre-processing separated the email body from the rest of the email and removed HTML tags if they were present. In addition, pre-processing for the intent analysis and emotion classification pipelines also included the splitting of email bodies into separate sets of words each with an associated set of labels such as relevant emotions or intent. 

\subsection{Implementation}

Recently, the advent of modern leading ML models 
has been propagated and bolstered largely thanks to the release of the ``Transformer'' by researchers at Google~\cite{vaswani2017attention}. 
As a consequence of its release, current leading models in terms of the universality of their applicability in the ML space such as T5~\cite{2020t5}, XLNet~\cite{yang2019xlnet}, and BERT~\cite{devlin-etal-2019-bert} all integrate the Transformer to varying degrees. Such universality means that they can be applied in a variety of different contexts, including emails, to obtain impressive results.

\textbf{Extractive Summarization}. 
We build a summarization algorithm that leverages recent transformer-based models: T5~\cite{2020t5}, XLNet~\cite{yang2019xlnet}, and BERT~\cite{devlin-etal-2019-bert}. T5 has produced the most promising results and, as a result, it has been adopted for this task. T5 has the advantage that it does not need to be fine-tuned for individual tasks by design. As a result, the implementation of the task and purpose with this model has been somewhat trivial. Whereas the utility of achieving the above may be evident, the precise manner in which content is to be summarized in an ideal scenario is of course open to interpretation since there are no objective criteria. As a result, fractional limits have been set with regards to the word length of a generated summary for any given email for examination in addition to hard-set limits such as 25, 50, or 100 words. Closer examination and some metric such as cross-validation is potentially needed to decide on how to best ascertain what word limit the summary should have - through admittedly looser criteria such as subjectively judging grammar, readability, and the ratio of summary length to essential context, the most promising result has been to set the maximal length of a summary to be a fraction of one fifth of the length of the original email.

\begin{figure}[h]
        \includegraphics[width=1\columnwidth]{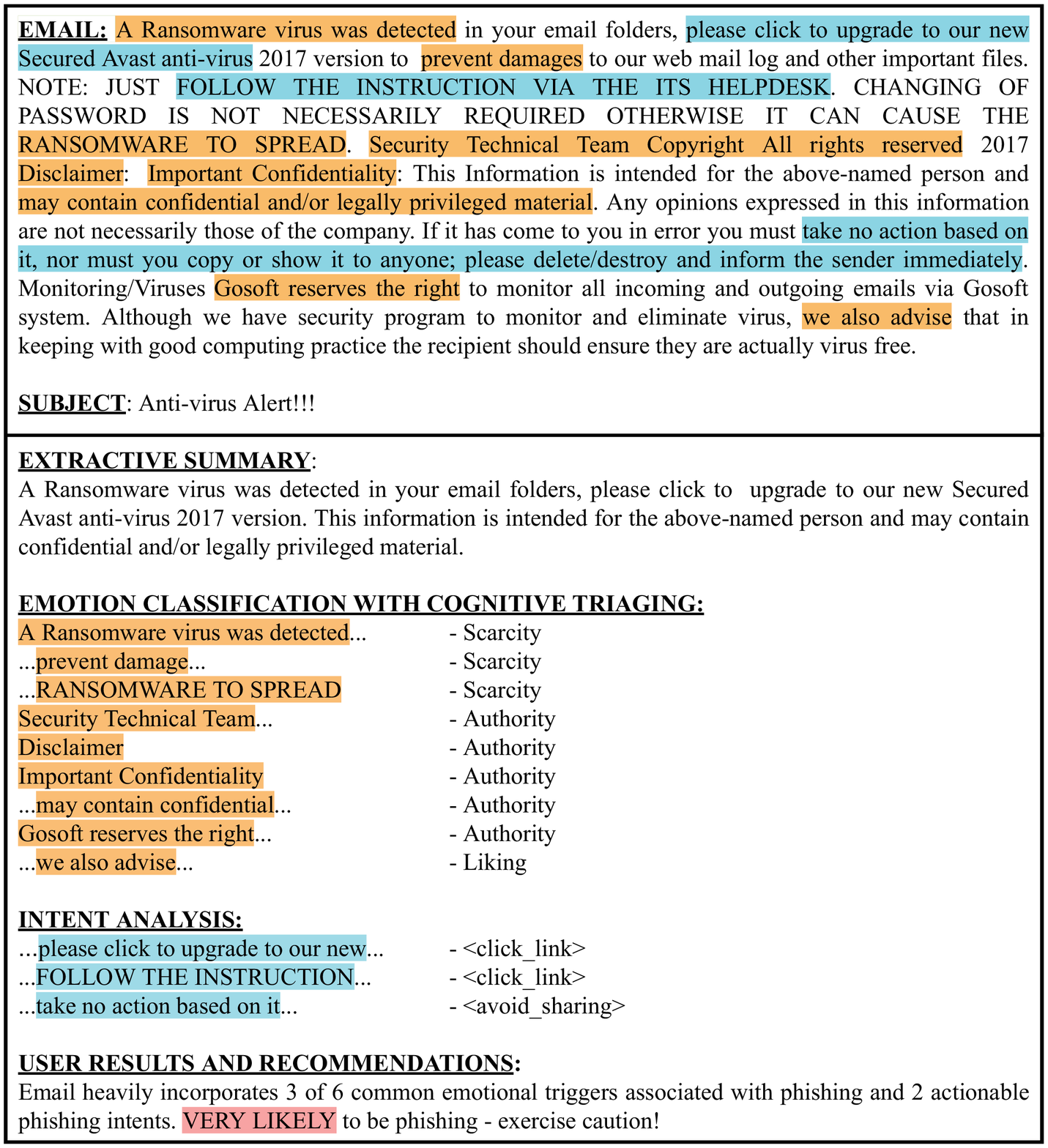}
        \caption{\label{summary.png}An example of a phishing email, results from the three pipelines, and the final system-generated user results.
        }
        \label{fig:extractedInfo}
\end{figure}

\textbf{Emotion Classification}. 
We first curated a random sample of sentences from 500 emails to be labelled under seven classes, with the additional class ``None'' represenitng that no given quality was present. This labelled set was then used to train the T5 and BERT models such that an individual set of words from any given email could have its most likely prevalent qualities identified by analyzing logit score outputs from the models implemented.  The overall assessment of the sentences and sets of words in the emails was then used to build a combined probability score that the email has a ``spike'' in one or more of the six cognitive triaging qualities - indicating a likely attempt to phish. Combined with assessments from the other branches the intent is to supply a user with all the prospective information they may need to make a rational and informed assessment themselves given what emotions are present and what they are trying to elicit in the context of the email - information generated with the help of this branch. 

\textbf{Intent Analysis.} The results have shown that the most promising models to implement this branch have been T5 and BERT, with T5 edging out BERT slightly with regards to accuracy and loss. To categorize common phishing intents that were present in emails, only short and apt descriptions have been added as tags to sets of word - such as ``click link'' or ``download file''. We randomly selected 500 emails from our datasets and conducted tagging manually. 
We then trained T5 to recognize malicious intents in the sentences of any email, with the objective of ``cherry-picking'' phishing intent in new emails users should pay particular attention to if other indicators such as cognitive triaging or the summary are suspicious. Figure~\ref{fig:extractedInfo} shows examples of intent evocations. We can note that the generated, concise summary contains the triggers and intent evocations at a much higher density, as we have suspected given our description in Section 2. By informing the user of the threshold criteria by which the tags are generated and the prospective danger of there being too many, the ambition is to achieve our set out objective of keeping the user alert. Experimentation at a larger scale is the next planned step to optimize this branch further.


\section{Concluding Remarks}
In this work, we devised a human-centric notification mechanism that extracts prospective psychological triggers, possible malicious intent, and a representative summary from emails. We then present the above in a meaningful way to the user for better decision-making and to elevate their learning of continuously evolving phishing patterns. Further examination with user studies and objective experimental analysis at a larger scale will give more insight into the effectiveness of this methodology. A concertized metric of trigger and intent density as a fraction of the total email length can also be examined to determine the metrics' correlation with whether an email is ``phishy'' or not.


\bibliographystyle{ACM-Reference-Format}
\bibliography{ref}


\end{document}